\documentclass[12pt]{article}

\usepackage{scicite}

\usepackage{times}
\usepackage[pdftex]{graphicx}
\usepackage{floatpag} %removes pagenumbers on Figure pages

\newenvironment{sciabstract}{%
\begin{quote} \bf}
{\end{quote}}

\newcounter{lastnote}
\newenvironment{scilastnote}{%
\setcounter{lastnote}{\value{enumiv}}%
\addtocounter{lastnote}{+1}%
\begin{list}%
{\arabic{lastnote}.}
{\setlength{\leftmargin}{.22in}}
{\setlength{\labelsep}{.5em}}}
{\end{list}}

\title{Zen and the Science of Pattern Identification: \\ An Inquiry into Bayesian Skepticism}

\author
{Deborah A. Striegel,$^{1}$ Damian Wojtowicz,$^{2}$ Teresa M. Przytycka,$^{2}$ Vipul Periwal$^{1\ast}$\\
\\
\normalsize{$^{1}$Laboratory of Biological Modeling, NIDDK, }\\
\normalsize{$^{2}$Computational Biology Branch, NCBI, NLM, }\\
\normalsize{The National Institutes of Health, Bethesda MD 20892, USA}\\
\\
\normalsize{$^\ast$To whom correspondence should be addressed; E-mail:vipulp@mail.nih.gov.}
}

\date{}

\begin{document} 

% Double-space the manuscript.

\baselineskip24pt

% Make the title.

\maketitle

% Place your abstract within the special {sciabstract} environment.

\begin{sciabstract}
Finding patterns in data is one of the most challenging open questions in information science. The number of possible relationships scales combinatorially with the size of the dataset, overwhelming the exponential increase in availability of computational resources. Physical insights have been instrumental in developing efficient computational heuristics. Using quantum field theory methods and rethinking three centuries of Bayesian inference, we formulated the problem in terms of finding landscapes of patterns and solved this problem exactly. The generality of our calculus is illustrated by applying it to handwritten digit images and to finding structural features in proteins from sequence alignments without any presumptions about model priors suited to specific datasets. Landscapes of patterns can be uncovered on a desktop computer in minutes.\end{sciabstract}

\clearpage

% In setting up this template for *Science* papers, we've used both
% the \section* command and the \paragraph* command for topical
% divisions.  Which you use will of course depend on the type of paper
% you're writing.  Review Articles tend to have displayed headings, for
% which \section* is more appropriate; Research Articles, when they have
% formal topical divisions at all, tend to signal them with bold text
% that runs into the paragraph, for which \paragraph* is the right
% choice.  Either way, use the asterisk (*) modifier, as shown, to
% suppress numbering.

%\section*{Introduction}

Technological advances have made large-scale data collection and storage increasingly practical. The problems we face today can often be reduced to deducing a pattern of relationships in data, in science and elsewhere. However,  super-exponential combinatorial growth in the number of possible relationships as the size of a datum increases hampers our ability to find relationships in large datasets. Thus, the development of new techniques for the discovery of relationships is critical to progress. 

Here we propose a framework based on physical heuristics to overcome this barrier to finding the set of fundamental interactions in the data, which we call a pattern. Physical intuitions in computation have an illustrious history. Thirty years ago, simulated annealing revolutionized optimization[1]. More recently, quantum computation has taken center stage[2]. Statistical physics has been applied to data analysis but usually in the form of approximations[3].  The data itself has unknown relationships and approximations such as high temperature expansions, weak-coupling expansions, mean field theory, and similar techniques[5] introduce further uncontrolled uncertainties. 

We sought an exact solution for finding patterns in data without presuming anything about the presence of patterns. Without any loss of generality, we consider datasets consisting of families of sequences of symbols and model such families with the traditional weight matrices where $p_{iB}$  is the probability of symbol $B$ at  sequence position $i.$ Datasets that contain a pattern have a  non-random weight matrix.   Specifically, we wanted to directly compute the relationships between sequence positions and the common deformations that relate different sequences in a family, which we define as the landscape of the pattern.  We developed  an approach, Bayesian Skepticism (BS), grounded in quantum field theory[4] that leads to an exact solution of such pattern landscapes. Our derivation stands on three conceptual legs: 
\begin{enumerate} 
\item A skeptical approach to modeling the data with a model prior probability distribution $\exp(MK(p|M))$ favoring random models. This is analogous to a null hypothesis, but with a continuous degree of skepticism $M$ in this assumption. This analytic continuation is similar to dimensional regularization in quantum field theory[4].
\item A reversal of Bayesian model selection: For a given random model, is there a weighting of data points that would make data frequencies consistent with model probabilities? 
\item A method that solves the maximization of model and data likelihoods exactly,  giving the weight of each datum and simultaneously uncovering the most likely model for the dataset. 
\end{enumerate}
The end result of our derivation[19] is a simple iteration of elementary algebraic steps: With $M\times K(p|M)$ the skeptical prior log probability of $p_{iA},$ the  observed weighted frequencies are updated as
\begin{equation}
f_{iA}^{(n+1)} \leftarrow \sum_{\rm sequences} \sigma_{i,A}({\rm sequence})\exp\bigg(M\sum_{j,B} \sigma_{j,B}({\rm sequence})[\partial K/\partial p_{jB}|_{p=f^{(n)}}-{\rm const.}]\bigg)
\end{equation}
where $(n)$ is the iteration count, and $\sigma_{i,A}$(sequence) is 1 if the sequence has symbol $A$ at position $i$ and 0 otherwise, and the constant ensures that the weights over all sequences sum to 1. Since we have been able to express the iterative update using simple algebra, it is easy to find the fixed point[19]. We can then find the  probability that a given string of symbols belongs to this family[19], and thus find the allowed deformations of the family. These are the correlated symbol probability changes such that the  probability of the altered model does not change significantly from that of the fixed point model. These deformations are the flat directions in the pattern landscape. 
 
We applied these concepts to the MNIST hand-written digit database[6]. This database consists of 60,000 gray-scale images of numbers $0$ through $9.$ We reduced these $28\times 28$ pixel format images to an unordered sequence of symbols by associating symbols with each of the 16 possible combinations of black and white pixels in a $2\times 2$ square (Fig. 1A). Each digit was thus associated with a family of about 6000 strings of symbols, each string of length 729 symbols (Fig. 1B,C) Our skepticism assumption, that the symbols occur randomly, suggests that the prior probability of a model of such sequences should be the generalization of the binomial coin-tossing probability distribution to a 16-sided coin toss independently for each position. The multinomial distribution at each position $i,$
\begin{equation}
\exp(MK(p_{iA}|M,q_A) )= M! \prod_C {q_C^{Mp_{iC}}/{(Mp_{iC})!}} 
\end{equation}
gives the probability that the 16 symbols $A$ will appear  $M\times p_{iA}$ times at position $i$ in a series of $M$ coin-tosses. We analytically continued this formula to small non-integer positive values of $M$ using the properties of the Gamma function. Typically, we will set  $M=0.005$ in this paper, and keep only sequence positions with less than 95\%\ conservation.  Qualitative results do not depend on either $M$ changed to 0.007 or conservation cutoff changed to 99\% [19].  $q_A$ is the frequency of symbol $A$ over all positions in the sequences corresponding to a digit.

The fixed-point of the iteration in eq.~1 can be used to find the flat directions. Changing the model along a flat direction induces entropy changes at each position in the sequence. By visualizing this entropy change[19], we observe that some positions in the sequence become more specialized and lose entropy and others gain entropy as the probabilities change along a flat direction. The three flattest directions for each digit are shown in Fig. 1D, and have obvious geometric meanings. 
One can deduce that most people are right-handed by noting the slant towards the right. These flat directions are global in character though the analysis made no use of the ordering of the sequence or of continuity. They are correlated changes in symbol frequencies that are consistent with the functional constraints of representing the digit.

The mutual information between sequence positions, computed from the pattern landscape effective potential by simultaneous perturbing symbol frequencies at two different positions, could recover structural connections between different sequence positions, enabling reconstruction of  the original image much like a jig-saw puzzle. We evaluated pair-wise interaction information (PII, defined in [19]) between all pairs of sequence positions and then restricted to positions with high enough symbol entropy. The entire digit can be recovered and shows the presence of long-range interactions that encode functional relations between non-adjacent positions (Fig. 2). Replacing the symbol entropy with other measures of positional information did not change these results. 

We turn now to sequences of symbols that do come with a particular order, the amino-acid sequences of families of proteins.  First, we show how pattern landscapes can be used to distinguish functionally different subclasses of homologous proteins. The NTPase {\it Maf\_Ham1} domain is shared by two subfamilies of proteins with a structurally similar nucleotide binding cleft  but altered conserved residue location and nature[15]. The  landscape computed from a subset of Subfamily 1 assigned a lower  probability to sequences of Subfamily 2 compared to a test set of Subfamily 1(Fig. 3A) and vice versa. Similar results are found in the case of digits.

Next, we demonstrate a relation between pattern landscapes and structural properties. The intimin/invasin family of virulence factors are produced by Gram-negative bacteria. The structure of these proteins was recently solved[16], and we computed the  landscape from 146 aligned sequences.  The actual landscape is much flatter than the landscape of column-wise randomly permuted sequences (Fig. 3B), as a small eigenvalue corresponds to a larger radius of curvature around the most likely model. Following steps analogous to the digit example, we computed the entropy changes induced by the most significant deformation (Fig. 3 C-E), revealing that this deformation corresponds to specializing either the side chains projecting towards the outside of the barrel or those projecting towards the inside. Investigating PII between sequence positions (Fig. 3F) as we lower the threshold for PII, we find disconnected sets of clusters that coalesce into a connected graph (Fig. 3G). We mapped this minimal single cluster of PII interacting residues on the structure revealing juxtaposition suggesting allostery as well as predicting longer range interactions (Fig. 3H). 

More subtle structural features in proteins can also be detected in the landscapes of sequence alignments. The dihydrofolate reductase ({\it DHFR}) enzyme has been studied for its conformational changes related to catalytic activity[18]. We computed PII for the sequences[20](Fig. 4E) and the deformations in the landscape. Fig. 4 B-D shows the structural elements that are detected by these deformations. The interacting cluster for {\it DHFR} deduced from PII is the whole protein (Fig. 4F) with the strongest interactions around the folate binding pocket.
[13] derived a residue-residue interaction matrix by computing side-chain folding-unfolding dynamics for the E. coli {\it DHFR} protein. This matrix is similar to the PII matrix (Fig. 4E) computed from sequences produced by evolutionary dynamics indicating that evolutionary history has tested side-chain compatibility.  

A pioneering approach to detecting structural features in {\it DHFR} and some other proteins from sequence data, based on random matrix theory[7,8,9] and analogies with finance, weights a binary approximation to the residue-residue correlation matrix with the derivative of the large $M$ limit of the binomial distribution. Eigenvectors of the absolute values of this weighted matrix have been associated with allosteric interactions[8]. The features we find capture regions of structural variation consistent with function as shown by mapping them to known protein structures (Fig. 4B-D,G) and the positioning suggests allostery as well. 

Explicit statistical physics-type energy models of residue interactions that have been extensively studied in various approximations[10,11,12]. These works compute contact maps for proteins with the help of additional secondary structure predictions. The major distinction is that BS is exact and has a consistent expansion to all orders in interactions[19], with a geometric interpretation in terms of the curvature of the landscape around the most likely model. The use of a skeptical model prior in BS is the key to these properties. While high PII values are likely to be associated with residue pairs that are spatially close, tertiary structure proximity is not necessarily the case. In the two-dimensional digit examples (Fig. 2), PII links between positions with high enough symbol entropy are geometrically meaningful but do not imply two-dimensional proximity. Understanding protein structure through clusters in graphs has been considered[14] though the graphs had a different origin.

Our method is applicable to a wide spectrum of dataset sizes. We used about 6000 sequences for each digit, but for the intimin/invasin and the two subfamilies of the {\it Maf\_Ham1} proteins we computed  landscapes with only 146, 67 and 70 sequences, respectively. Our results are robust to choices of sequences. For {\it DHFR}, while we used 1382 sequences from OrthoDB[21] without any curation in our analysis, we also obtained the 417 sequences used in [7,8] and found that, for the positions present in both alignments, there were no striking differences between PII computed from the two alignments. 

To compare and contrast BS with common Bayesian approaches, Expectation Maximization (EM) [22] is an iterative algorithm for finding the maximum likelihood parameters of model or latent variables that maximize the probability of the dataset. EM requires no model prior for its interleaving of expectation and maximization steps. Maximum Entropy (MaxEnt) model priors attempt to be maximally unbiased or uninformative. BS has no maximization step in its iteration and cannot work without the skeptical model prior that explicitly precludes correlations, thus going beyond MaxEnt. Crucially, BS is self-limiting in how large the skepticism parameter $M$ can become because, by construction, the pattern landscape is convex[19]. A smaller dataset begets a more restricted landscape. This is evidenced in the reduced radius of curvature associated with larger eigenvalues (Fig. 4A) for {\it DHFR} for the alignment used in [7,8] compared to [20].

In summary, BS provides a simple exact solution to computing patterns in data. It is conceptually transparent with an explicit derivation and a venerable interpretation[4] in terms of a landscape of models. The skeptical approach introduced here of assuming a model prior that explicitly favors the lack of a pattern as a tool to deduce the actual pattern landscape from the data is broadly applicable, and completely side-steps questions of the biological relevance of model priors. We have shown that it can be used to compute interactions between sequence positions exactly and effectively. Of particular interest is the holistic character of the deformations we find: No single position can capture the `8-ness' of a sequence of symbols associated with the digit $8$, yet symbol entropy changes in flat directions capture the slant differences and other flourishes in handwriting. Molecular dynamics (MD) simulations of protein activity are hampered by the computational difficulty of sampling transient high-energy states. These motions are predicted to occur in micro-second to milli-second ranges, can be functionally important, and are difficult to model with atomic-scale MD[17]. The discovery of structural elements in protein structures from sequence alignments using BS may help overcome some of these difficulties in MD by suggesting structural elements capable of independent motion[17]. Further, BS can directly extract networks of interactions from data exploring the role of heterogeneity and variation in biological systems[23]. The protean character of BS  is well-suited to diverse applications in science.

%In this file, we present some tips and sample mark-up to assure your
%\LaTeX\ file of the smoothest possible journey from review manuscript
%to published {\it Science\/} paper.  We focus here particularly on
%issues related to style files, citation, and math, tables, and
%figures, as those tend to be the biggest sticking points.  Please use
%the source file for this document, \texttt{scifile.tex}, as a template
%for your manuscript, cutting and pasting your content into the file at
%the appropriate places.
%
%{\it Science\/}'s publication workflow relies on Microsoft Word97.  To
%translate \LaTeX\ files into Word97, we use an intermediate MS-DOS
%routine \cite{tth} that converts the \TeX\ source into HTML\@.  The
%routine is generally robust, but it works best if the source document
%is clean \LaTeX\ without a significant freight of local macros or
%\texttt{.sty} files.  Use of the source file \texttt{scifile.tex} as a
%template, and calling {\it only\/} the \texttt{.sty} and \texttt{.bst}
%files specifically mentioned here, will generate a manuscript that
%should be eminently reviewable, and yet will allow your paper to
%proceed quickly into our production flow upon acceptance \cite{use2e}.
%
{\bf Acknowledgements:} K. Reynolds and R. Ranganathan shared alignments of proteins. This work was supported by the Intramural Research Program of the National Institutes of Health, NIDDK and NLM.

%\section*{Formatting Citations}
%
%Citations can be handled in one of three ways.  The most
%straightforward (albeit labor-intensive) would be to hardwire your
%citations into your \LaTeX\ source, as you would if you were using an
%ordinary word processor.  Thus, your code might look something like
%this:
%
%
%\begin{quote}
%\begin{verbatim}
%However, this record of the solar nebula may have been
%partly erased by the complex history of the meteorite
%parent bodies, which includes collision-induced shock,
%thermal metamorphism, and aqueous alteration
%({\it 1, 2, 5--7\/}).
%\end{verbatim}
%\end{quote}

%\noindent Compiled, the last two lines of the code above, of course, would give notecalls in {\it Science\/} style:
%
%\begin{quote}
%\ldots thermal metamorphism, and aqueous alteration ({\it 1, 2, 5--7\/}).
%\end{quote}
%
%Under the same logic, the author could set up his or her reference list as a simple enumeration,
%
%\begin{quote}
%\begin{verbatim}
\clearpage
{\bf References and Notes}

\begin{enumerate}
\item S. Kirkpatrick, C.D. Gelatt, M.P. Vecchi, {\it Science} {\bf 220}, 671-680 (1983).
\item P. Shor, {\it SIAM J. Comput.} {\bf 26}, 1484-1509 (1997).
\item S.V. Buldyrev, N.V. Dokholyan, A.L. Goldberger, S. Havlin, C.-K. Peng, H.E. Stanley and G.M. Viswanathan, {\it Physica A} {\bf 249}, 430-438 (1998).
\item J. Zinn-Justin, {\it Quantum Field Theory and Critical Phenomena} {(Oxford Univ. Press, New York, 2002).}
\item L.D. Landau and E.M. Lifschitz, {\it Statistical Physics} (Pergamon Press, Oxford, 1980).
\item Y. LeCun, L. Bottou, Y. Bengio, and P. Haffner, {\it Proc. of the IEEE} {\bf 86}, 2278-2324 (1998); MNIST database available at http://yann.lecun.com/exdb/mnist/
\item N. Halabi, O.  Rivoire, S. Leibler, and R. Ranganathan, {\it Cell} {\bf 138}, 774-786  (2009).
\item K.A. Reynolds, R.N. McLaughlin, and R. Ranganathan, {\it Cell} {\bf 147}, 1564-1575 (2011). 
\item R.N. McLaughlin, F.J. Poelwijk, A. Raman, W.S. Gosal, and R. Ranganathan,  {\it Nature} {\bf 138}, 138-142 (2012).
\item D.S. Marks, T.A. Hopf, and C. Sander, {\it Nat. Biotechnol.} {\bf 30}, 1072-80 (2012).
\item F. Morcos, A. Pagnani, B. Lunt, A. Bertolino, D.S. Marks, C. Sander, R. Zecchina, J.N. Onuchic, T. Hwa, and M. Weigt,
{\it Proc Natl Acad Sci U S A}  {\bf 108}, E1293-301 (2011).
\item D.S. Marks, L.J. Colwell, R. Sheridan, T.A. Hopf, A. Pagnani, R. Zecchina, and C. Sander,  {\it PLoS ONE} {\bf 6}, e28766  (2011).
\item H. Pan, J.C. Lee, and V.J.  Hilser, {\it Proc Natl Acad Sci U S A} {\bf 97}, 12020-5 (2000).
\item A. Sukhwal, M. Bhattacharyya and S. Vishveshwara, {\it Acta Cryst. D} {\bf 67}, 429-439 (2011).
\item S. Chakrabarti, S.H. Bryant, and A.R. Panchencko, {\it J. Mol. Biol.} {\bf 373}, 801-810 (2007). 
\item J.W. Fairman, N. Dautin, D. Wojtowicz, W. Liu, N. Noinaj,T.J. Barnard, E. Udho, T.M. Przytycka, V. Cherezov and S.K. Buchanan, {\it Structure} {\bf 20}, 1233-1243 (2012).
\item B.K. Ho and D.A. Agard, {\it PLoS Comp. Bio.} {\bf 5}, e1000343 (2009).
\item J.R. Schnell, H.J. Dyson and P.E. Wright, {\it Annu. Rev. Biophys. Biomol. Struct.} {\bf 33}, 119-140 (2004).
\item The derivation and supplementary supporting figures are available as supplementary text S1.
\item DHFR alignment from OrthoDB, supplementary text S2.
\item R.M. Waterhouse, E.M. Zdobnov, F. Tegenfeldt, J. Li, and E.V. Kriventseva, {\it Nucl. Acids Res.} {\bf 41}, D358-365 (2013); http://cegg.unige.ch/orthodb6.
\item A.P. Dempster, N.M. Laird and D.B. Rubin, {\it J. Royal Stat. Soc. Ser. B} {\bf 39}, 1-38 (1977).
\item S.J. Altschuler and L.F. Wu, {\it Cell} {\bf 141}, 559-563 (2010).

%\item G. Gamow, {\it The Constitution of Atomic Nuclei
%and Radioactivity\/} (Oxford Univ. Press, New York, 1931).
%\item W. Heisenberg and W. Pauli, {\it Zeitschr.\ f.\ 
%Physik\/} {\bf 56}, 1 (1929).
\end{enumerate}
\clearpage
Figure 1. Sequence representation of hand-written digits. A. The sixteen possible configurations of black and white pixels that correspond to distinct letters. B. The grid of 28$\times$28 squares superimposed on a specific example of the digit 5. C. The linear sequence of 729 letters obtained in this manner is shown arranged in a $27\times27$ square to exhibit the digit. Overlapping $2\times 2$ squares were mapped to the letters associated with the configurations in A. D. Symbol entropy changes in the three flattest directions in the pattern landscapes of all 10 digits, relative to the most likely model, with 25 example images for each digit.

Figure 2. Interactions between sequence positions for digits. First column: PII between position pairs represented as a heat map for each digit. Second column: PII values vs. distance in the two-dimensional image. All black PII values (PII $>-0.77$) are shown along with a sampling of the other regions (1\% and 0.5\% of PII values from blue and brown, and orange and red regions, respectively) Inset: Histogram of all PII values (log (base 10) scale on $y$-axis). Third column: PII values plotted as edges on a graph on the digit image.  Edges correspond to points observed in column 2. The heat map background color scale represents symbol entropy at each position. Fourth column: Edges consist of the strongest PII-value pairs for each position that also have high symbol entropy. %PII values only between positions with high enough symbol entropy, computed by lowering a threshold PII cutoff until the graph has only one component. 
 Most long-range interactions are removed by this cutoff, with the remaining interactions %(proximal and long-range) 
 of clear geometric meaning. Fifth column: PII values above the symbol entropy threshold vs. two-dimensional distance. All black PII values (PII $>-0.77$) are shown along with a sampling of the other regions (1\% and 0.5\% of PII values from blue and brown, and orange and red regions, respectively) Inset: Histogram of all PII values above threshold (log (base 10) scale on $y$-axis).

Figure 3. Applications to {\it Maf$\_$Ham1} and intimin/invasin proteins. A. Upper panel: Probability of {\it Maf$\_$Ham1} sequences in the pattern landscape of Subfamily 1. Sequence index is plotted on the $x$-axis and $-\log$ Probability on the $y$-axis. The first 12 sequences are test sequences from Subfamily 1 not used in the computation of the pattern landscape and the rest are all the sequences in Subfamily 2. Only the linear term in the pattern probability $M\Gamma(p)$  between positions with above-median symbol entropy is used in this computation, consistent with the results shown in Fig. 2. Lower panel: Similar probabilities with the roles of Subfamily 1and Subfamily 2 reversed. $M=0.01$ for these probabilities. B. First 20 eigenvalues plotted in increasing order for intimin/invasin. The second eigenvalue is almost twice as large as the lowest eigenvalue, suggesting that the landscape has only one major flat direction. Inset shows eigenvalues from two instances of column-wise randomized intimin/invasin sequences showing much larger eigenvalues, and no corresponding isolated flattest direction. C,D,E. Front, cut-through and back views of the intimin/invasin beta barrel. Red positions are losing symbol entropy and blue positions are gaining entropy, relative to the most likely model (or vice-versa due to linearity). The symbol entropy changes are consistent with beta-barrel shear and this consistency is absent in column-wise randomized sequences[19]. F. PII heat map between all pairs of sequence positions. G. Graph of correlated residues with edges corresponding to PII values above a minimal threshold and nodes for only positions with high enough symbol entropy.  The minimal threshold PII cutoff is obtained by lowering a cutoff until the graph has only one component. Edges are colored according to PII value. H. The resulting correlated residues shown on the structure, colored according to their strongest PII link. As in the digit examples (Fig. 2),  the graph is not restricted to tertiary structure proximity.

Figure 4. Identification of structural features from flat directions for {\it DHFR}. A. Similar pattern of eigenvalues in ascending order for {\it DHFR}, for {\it DHFR} (metazoa, fungi, bacteria) sequences from OrthoDB[21] and for {\it DHFR} sequences used in [7] obtained from R. Rangananthan and K. Reynolds. Inset: Eigenvalues for column-wise randomizations are much higher than the eigenvalues for the actual sequences. B. The flattest direction in the {\it DHFR} pattern landscape picks out structural changes in the folate binding pocket. Symbol entropy increases (blue) in the folate binding pocket and decreases in the surface residues (red), or vice versa, relative to the most likely model. C: The strongest association of the second flat direction is with the G-H loop which plays a stabilizing role in interactions with the Met20 loop[18]. D: The junction between the $\alpha 4$ helix and the $\beta 6$ sheet is associated with the third flat direction. B,C,D Insets: Top to bottom: Specific structural changes in E. coli, chicken and human {\it DHFR} associated with these flat directions. E. PII heat map for {\it DHFR}. F.  Graph of correlated residues. For {\it DHFR}, all residues appear in the graph, consistent with known conformation changes  associated with ligand binding. Edges are colored according to PII value. G. {\it DHFR} structure residues colored according to their strongest PII link value. The strongest PII interaction is shown with side-chains.

\floatpagestyle{empty}

\begin{figure}[h]
	\centering
		\centerline{\includegraphics[angle=270,origin=c]{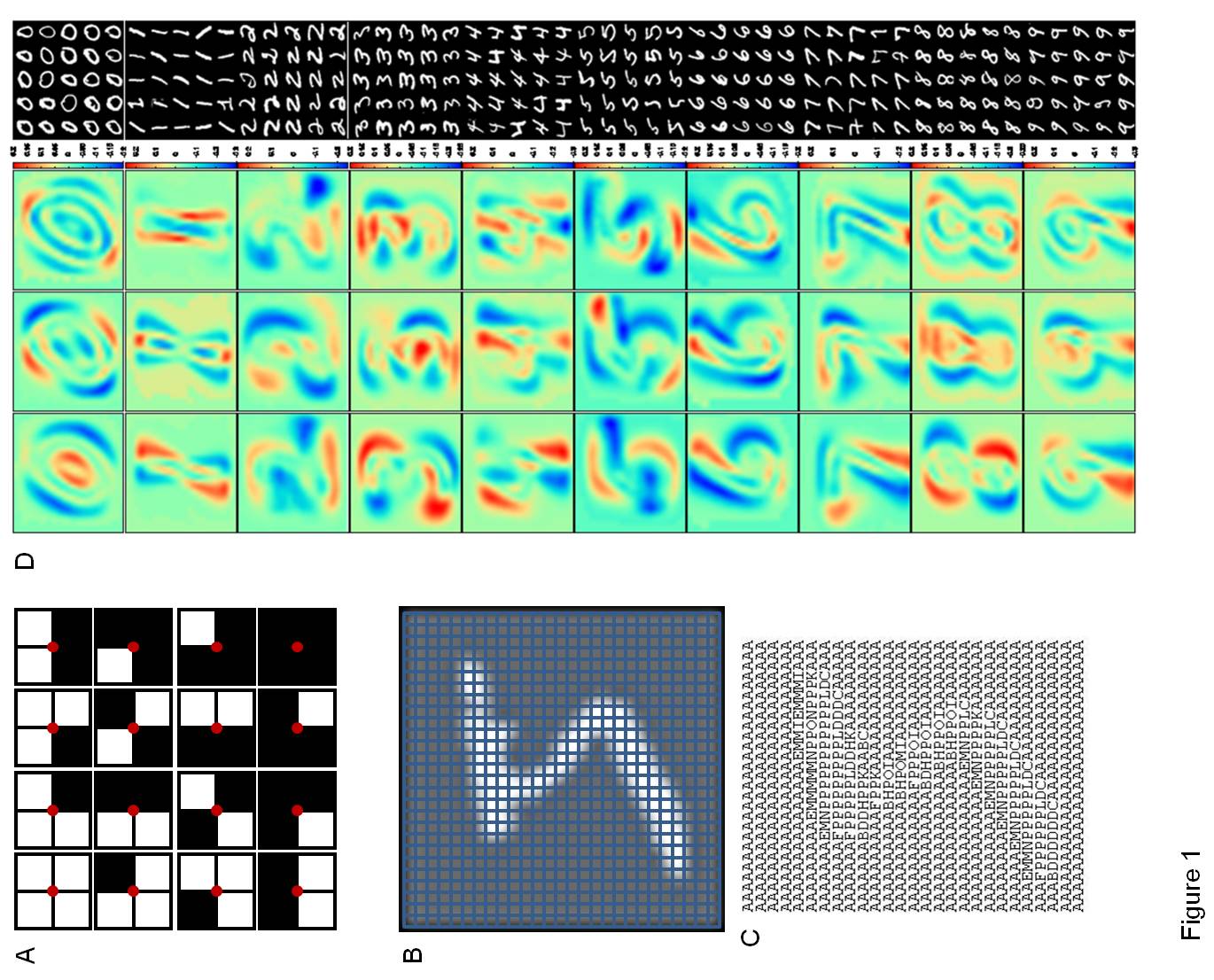}}
	\label{fig:Patterns_Fig1}
\end{figure}

%\hspace*{-1cm}

\begin{figure}[h]
	\centering
		\centerline{\includegraphics[height = 0.72\textheight,angle=270,origin=c]{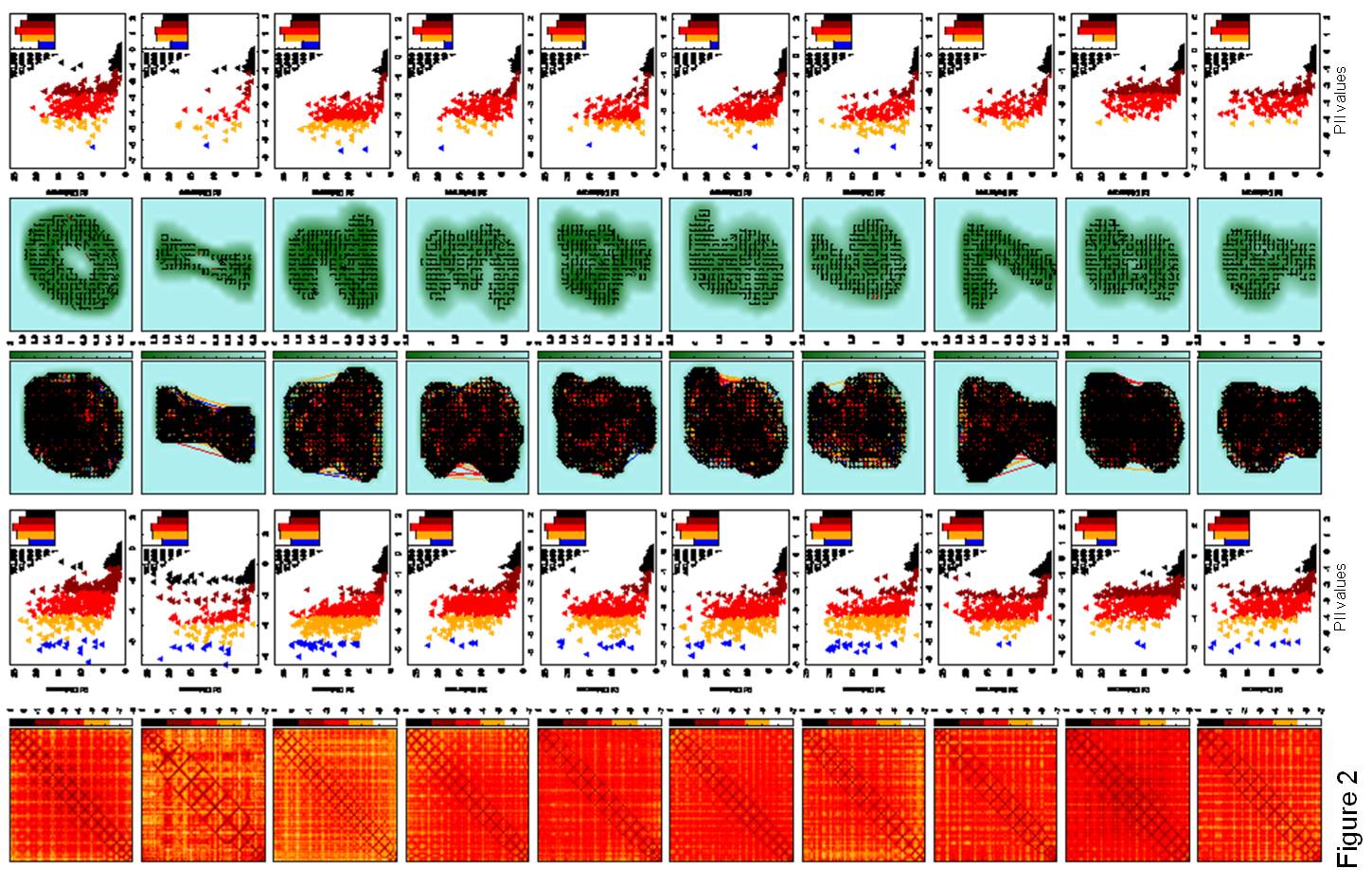}}
	\label{fig:Patterns_Fig2}
\end{figure}

\begin{figure}[h]
	\centering
		\centerline{\includegraphics[ angle=270,origin=c]{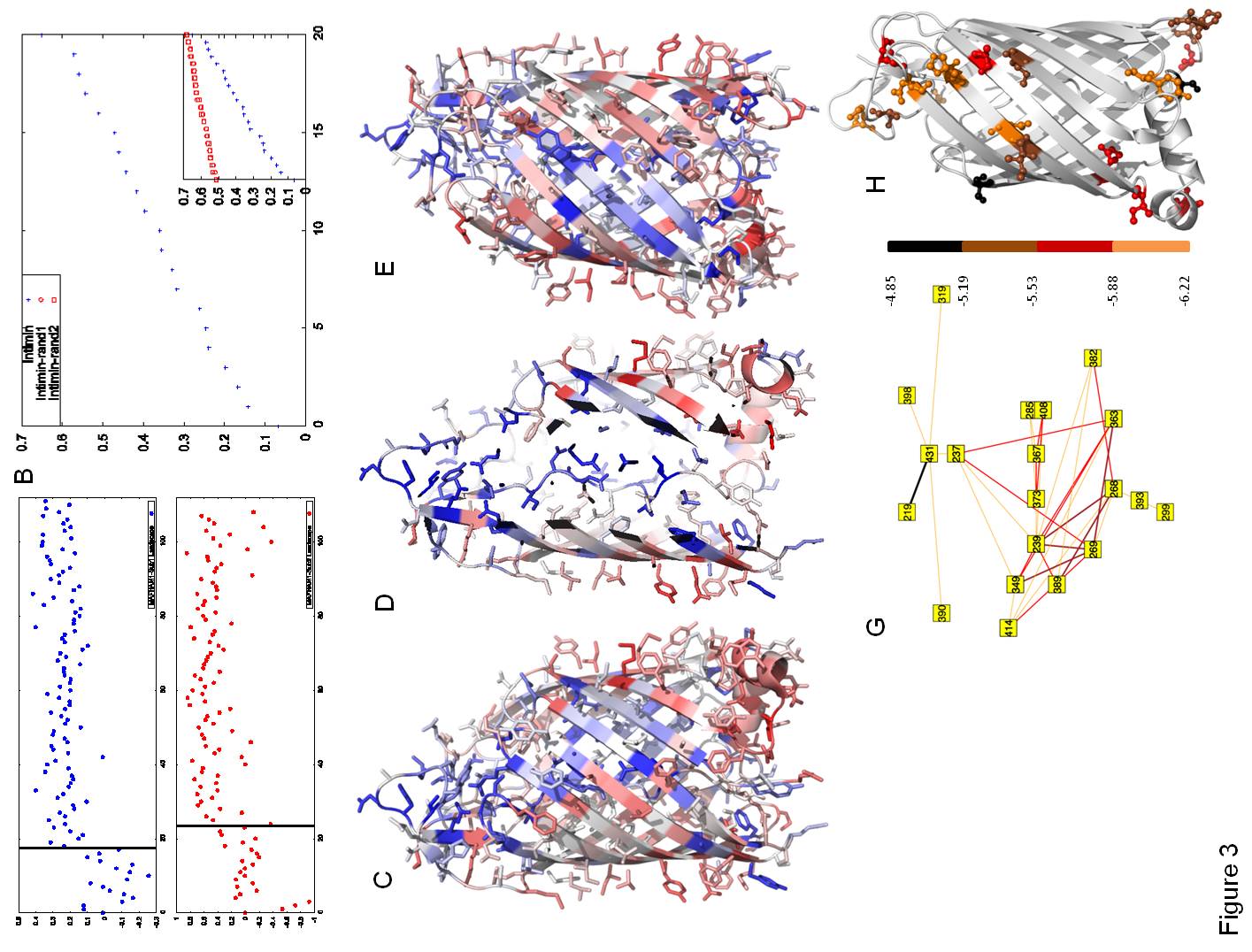}}
	\label{fig:Patterns_Fig3}
\end{figure}

\begin{figure}[h]
	\centering
		\centerline{\includegraphics[angle=270,origin=c]{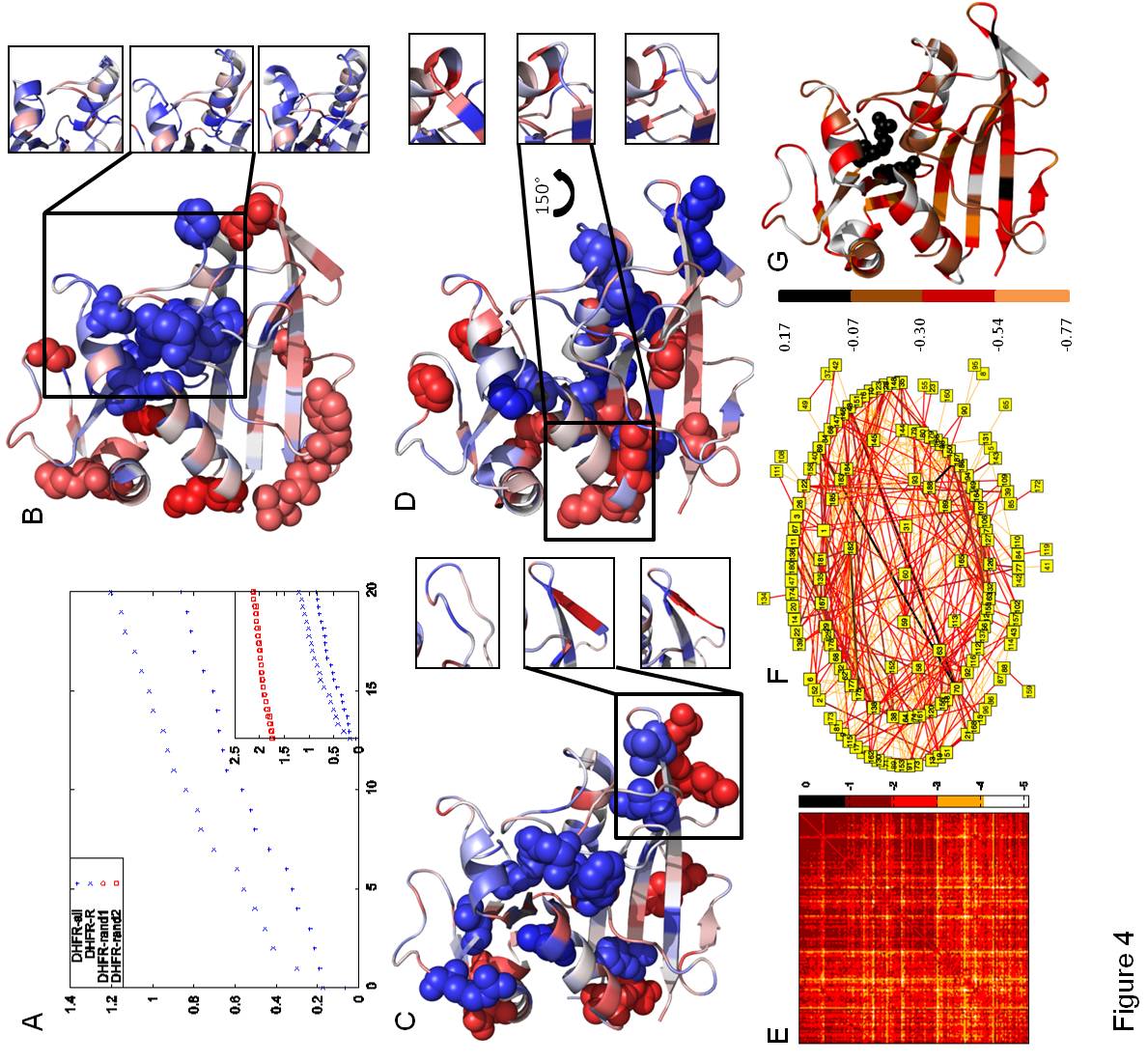}}
	\label{fig:Patterns_Fig4}
\end{figure}

%\end{verbatim}
%\end{quote}
\end{document}

% Double-space the manuscript.

\baselineskip24pt

% Make the title.

\maketitle 
\section*{Derivation}
Given a set of sequences of symbols $\{s\},$ we define a statistical sum with a symbol and position dependent weight $J_{iA}$ as follows:
 Let $N$ be the number of sequences, $G$ the number of different symbols and $R$ the number of positions in each sequence. The ordering implicit in writing the sequences as strings of symbols is not used in our analysis. Any permutation of the positions in the sequences will lead to the same result. Here $\sigma$(sequence) is a binary vector of length $R\times G$ with value 1 at position $(i,A)$ if symbol $A$ occurs at position $i$ in the sequence, and 0 otherwise. We define $W$ as the connected correlation function generating functional: 
 \begin{equation}
 \exp(MW(J) ) \equiv \sum_{\{S\}} \exp(MJ\cdot \sigma)
 \end{equation}
By taking derivatives with respect to $J,$ one computes connected normalized correlation functions in quantum field theory. We have introduced $M$ as a free parameter here that will be related to the skeptical multinomial prior in the latter part of the derivation.
The Legendre transform is a standard tool in transforming variables in the context of Newtonian dynamics relating Lagrangian to  Hamiltonian formulations of Newton's equations, and is also used in thermodynamics. The fundamental convexity properties of the Legendre transform should be kept in mind.
In the present context, the Legendre transform of $W$ is the effective potential $\Gamma$[4]:
 \begin{equation}
 W(J) + \Gamma(f) =  J \cdot f = \sum J_{iA} f_{iA}
 \end{equation}
 where $f_{iA}$ is the weighted frequency of symbol $A$ at position $i.$ As 
\begin{equation}
\langle \sigma_{iA} \rangle_J = {{\partial W}\big/ {\partial J}} = f_{iA}(J),
 \end{equation}
 the frequencies are the expectation values of $\sigma_{iA}.$
The effective potential defined in this manner $\Gamma$ is a function of $f_{iA},$ and the symmetry of the Legendre transform gives
\begin{equation}
{{\partial\Gamma(f)}\big/{\partial f_{iA}}} = J_{iA}(f) .
 \end{equation}
 Note that $\sum_A f_{iA} = 1$ so $J$ at each position is only determined up to an additive constant, which can be subtracted, allowing us to normalize $J$ by demanding $\sum_A J_{iA} =0.$
 $\exp(-M\Gamma(f))$ gives the probability of observing frequencies $f_{iA}$ by an appropriate choice of weights $\exp(MJ\cdot\sigma)$ for sequences. Equating model probabilities $p$ with observed weighted frequencies $f,$ we can ask for the model that is most likely given our skeptical multinomial prior
\begin{equation}
\hbox{\rm e}^{MK(p_{iB}|M,q_A)} = \bigg({M!/{\prod_A (Mp_{iA})!}} \bigg)\prod_A q_A^{Mp_{iA}} ,
\end{equation}
which explicitly assumes that there is no correlation between the symbols that appear at different positions in the sequence.
Following standard Bayesian inference, we want to maximize $K - \Gamma:$
\begin{equation}
{\partial K/{\partial p_{iA}}} - {\partial \Gamma/{\partial {p_{iA}}}} =0
 \end{equation}
 Combining eq.~4 and eq.~6 relating $J, \Gamma$ and $K,$ we find
\begin{equation}
J(p)=\partial K/\partial p 
 \end{equation}
 at the most likely model.
This is the crucial step because the solution of eq.~3 can now be written as an explicitly computable exact fixed point iteration:
\begin{equation}
f^{(n+1)} = \langle \sigma \rangle_{J = \partial K/\partial p\big|_{p=f^{(n)}}}.
 \end{equation}
 Once we find the fixed point values  $p_E,$ we know the exact weights of the sequences through knowing $J(p_E).$ This lets us use standard results on the Legendre transform to compute exact two and higher point connected correlation functions. Thus
\begin{equation}
\langle\sigma \sigma\rangle_{J(p_E)} = {\partial^2 W/{\partial J}^2} =  \left({\partial^2 \Gamma/{\partial p}^2}\right)^{-1} .
 \end{equation}
This evaluation of the second (and higher) derivatives of $\Gamma$ lets us expand the  probability $\Gamma$ in a Taylor series for any model $p_{iA}:$
\begin{equation}
\Gamma(p)= \Gamma(p_E) + (p-p_E){{\partial \Gamma}/{\partial p}}(p_E) + {1\over 2} (p-p_E)^2{{\partial^2\Gamma}/{\partial p^2}}(p_E) + \ldots
 \end{equation}
This expansion can be used to compute the probability of any symbol probability distribution. In particular, it can be used to compute the probability of any specific sequence, as a sequence corresponds to a probability distribution with frequencies taking values 0 or 1 at each position. 

To gain an intuition for the fixed point equation, recall that for small $x$ the Gamma function is given by 
\begin{equation}
\ln\Gamma (1+x) = 1-\gamma x+(\pi^2/12)x^2 + \ldots
\end{equation}
It follows that the fixed point values of $f_{iA}$ are the solution to
\begin{equation}
\sum_{s \in {\rm sequences}} (\sigma_{jB}(s) - f_{jB})\prod_{i,A} \bigg[q_A^M\exp\big( -\pi^2M^2f_{iA}/6\big)\bigg]^{\sigma_{iA}(s)}  = 0
\end{equation}
for each $j,B$ for small enough $M.$

We define the pairwise interaction information (PII) between two positions $i$ and $j$ in the sequence as the log mutual information computed from the probability of deviating from the fixed point model by altering the frequencies of symbols only at these two positions.
To be explicit,
\begin{equation}
PII(i,j) \equiv \ln \sum_{A,B} {\rm Pr}(iA,jB) \ln ({\rm Pr}(iA,jB)/{\rm Pr}(iA,j*){\rm Pr}(i*,jB))
 \end{equation}
 where 
 \begin{equation}
\ln {\rm Pr}(iA,jB) \propto MJ\cdot(\delta_{iA}+\delta_{jB}) + \delta_{iA}\delta_{jB}{{\partial^2\Gamma}/{\partial p^2}}(p_E)_{iA,jB}
 \end{equation}
 and $\delta_{iA}$ is a unit change in the probability of symbols at position $i,$ increasing the probability of symbol $A$ and decreasing the probability of all other symbols at that position appropriately to maintain unit probability. We normalize the probability ${\rm Pr}$ before computing PII.
 
The eigenvectors of the small eigenvalues of ${{\partial^2\Gamma}/{\partial p^2}}(p_E)$ correspond to flatter directions in the  landscape[4]. 
 We computed the entropy flow in the direction of any of these eigenvectors as follows: If $v_{iA}$ is an eigenvector of ${{\partial^2\Gamma}/{\partial p^2}}(p_E),$ we compute the derivative of the position-wise Shannon entropy in the direction of $v_{iA}$ as 
\begin{equation}
{\cal L}_{v}I_i = d/dt \bigg(-\sum_A [(p_{iA}+tv_{iA})/n_i(t)] \ln [(p_{iA}+tv_{iA}) /n_i(t)]\bigg)\bigg|_{t=0}, 
 \end{equation}
  where $n_i(t) = \sum_B (p_{iB}+tv_{iB})$ in order to keep probabilities normalized. 
  
The value of $M$ does not matter for qualitative features, see Supplementary Figures 2,5. The iteration does not converge and gives evidence of phase transition like behavior at large enough $M,$ with the entropy of the weights of the data sequences vanishing when only one specific datum has a weight approximately equal to unity and the other elements of the dataset have weights close to vanishing. There is also an example when the iteration reaches a limit cycle rather than a fixed point for $M$ large enough. $M=0$ is a singular point and none of our results apply. The convexity properties of the Legendre transform are responsible for the self-limiting features of the fixed-point iteration. 

\section*{Implementation}
We implemented this iteration and computation in C++. The matrix manipulations were performed using NumPy in the Enthought Python Distribution (www.enthought.com). The connected correlation function matrices have null eigenvalues corresponding to the obvious constraints on the frequencies $\sum_A f_{iA} =1$ for each position $i.$ These are trivially projected out and we inverted the matrices on the non-null subspace. When the number of sequences $N$ is less than $R\times (G-1)$  the non-zero eigenvalues are obviously reduced to $N.$ The number of non-zero eigenvalues cannot exceed the number of binary sequence vectors! For sequences that are closely related, we found that the numerically stable range of eigenvalues can be less than $N$ and is easily detected by inspection of the range of values of the matrix elements of the inverted matrix. 

The figures in the paper all consider sequence positions with less than 95\%\  conservation. Changing from 95\%\  to 99\%\  conservation did not change any qualitative features (e.g. Suppl. Fig. 3).
 
\section*{Supplementary Figures}
\begin{enumerate}
\item Symbol entropy changes in the three flattest directions for randomized digit sequences.
\item Intimin flattest direction symbol entropy changes at $M=0.007.$ A,B,C: Front, cut-through and back views.
\item Intimin flattest direction symbol entropy changes with positions with less than 99\% conservation. A,B,C: Front, cut-through and back views.
\item Two randomized intimin sequences: symbol entropy changes in flattest direction (A-C, randomization \#1 front, cut-through and back views; D-F, randomization \#2 identical views). Note that the $\beta$-sheet shear is no longer consistent with the entropy changes and the entropy changes are distributed between the inside and the outside of the barrel, in contrast to the actual intimin sequences which show a segregation (main text, Fig. 3C-E).
\item {\it DHFR} sequences: Comparison of symbol entropy changes for three flattest directions for $M=0.005$ (A-C) (also shown in main text, Fig. 4 B-D) and $M=0.007$ (D-F).
\item Randomized {\it DHFR} sequences: symbol entropy changes in three flattest directions for three different randomizations (\#1:A-C, \#2:D-F, \#3:G-I).
\item Comparison of symbol entropy changes for {\it DHFR} sequences used in [7] obtained from R. Rangananthan and K. Reynolds (A-C) and obtained from OrthoDB[21] (D-F).
\item Highest segment of PII values for {\it DHFR} sequences for ease of comparison to [13], showing segments highlighted in [13].
\end{enumerate}

\floatpagestyle{empty}

\begin{figure}[h]
	\centering
		\centerline{\includegraphics[height=0.5\textheight,angle=270,origin=c]{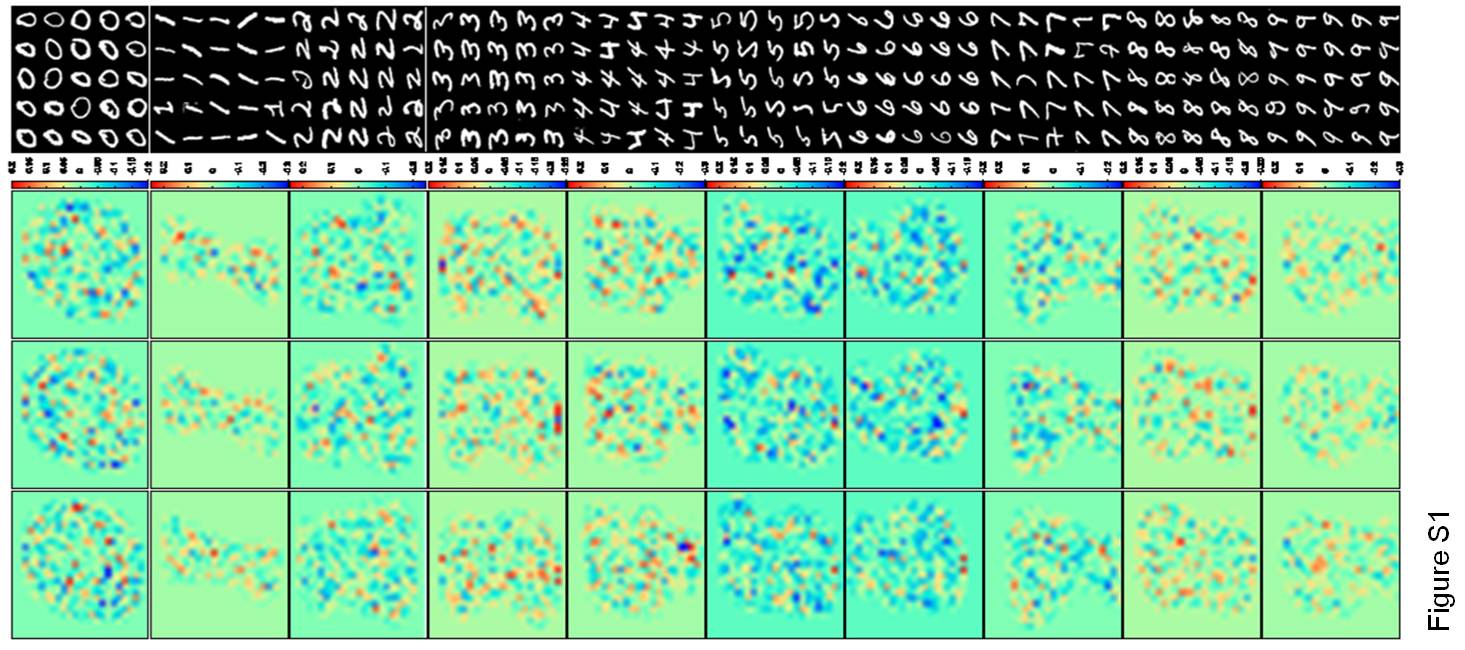}}
	\label{fig:Patterns_FigS1}
\end{figure}

%\hspace*{-1cm}

\begin{figure}[h]
	\centering
		\centerline{\includegraphics[angle=270,origin=c]{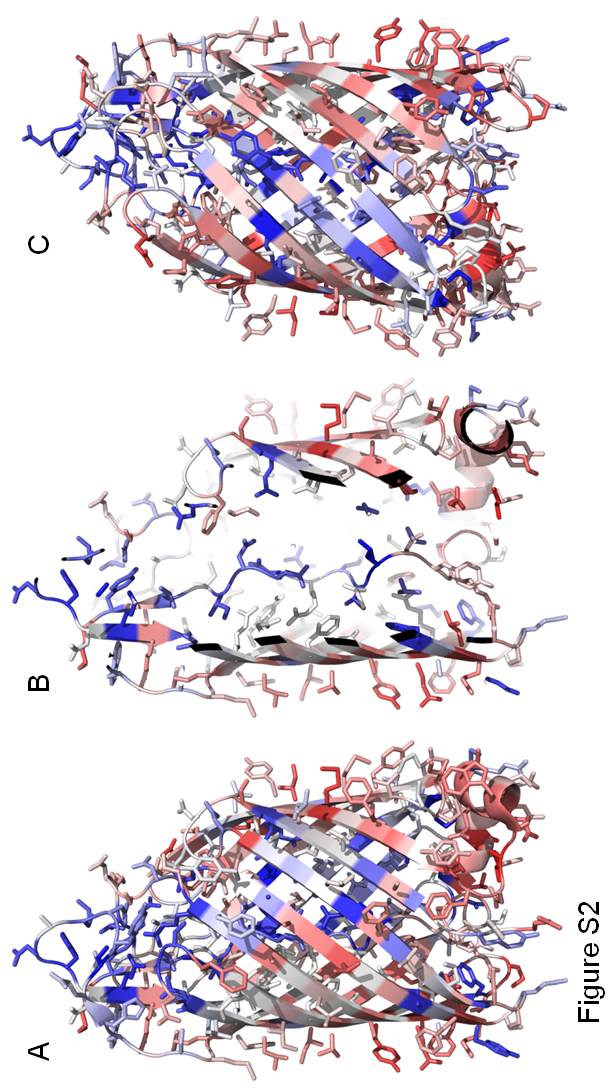}}
	\label{fig:Patterns_FigS2}
\end{figure}

\begin{figure}[h]
	\centering
		\centerline{\includegraphics[ angle=270,origin=c]{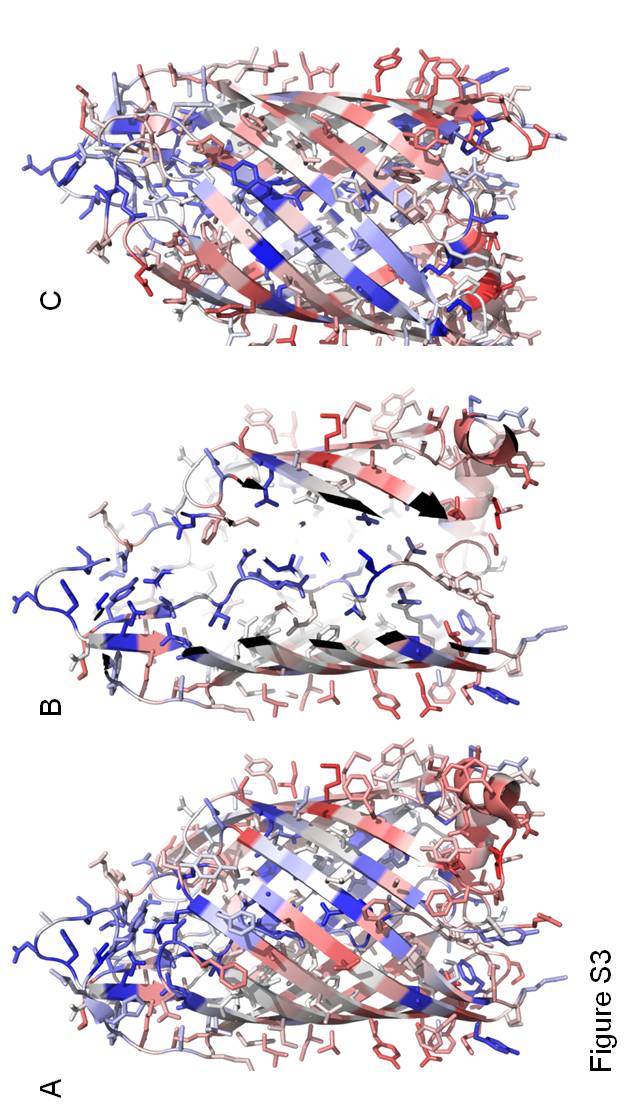}}
	\label{fig:Patterns_FigS3}
\end{figure}

\begin{figure}[h]
	\centering
		\centerline{\includegraphics[angle=270,origin=c]{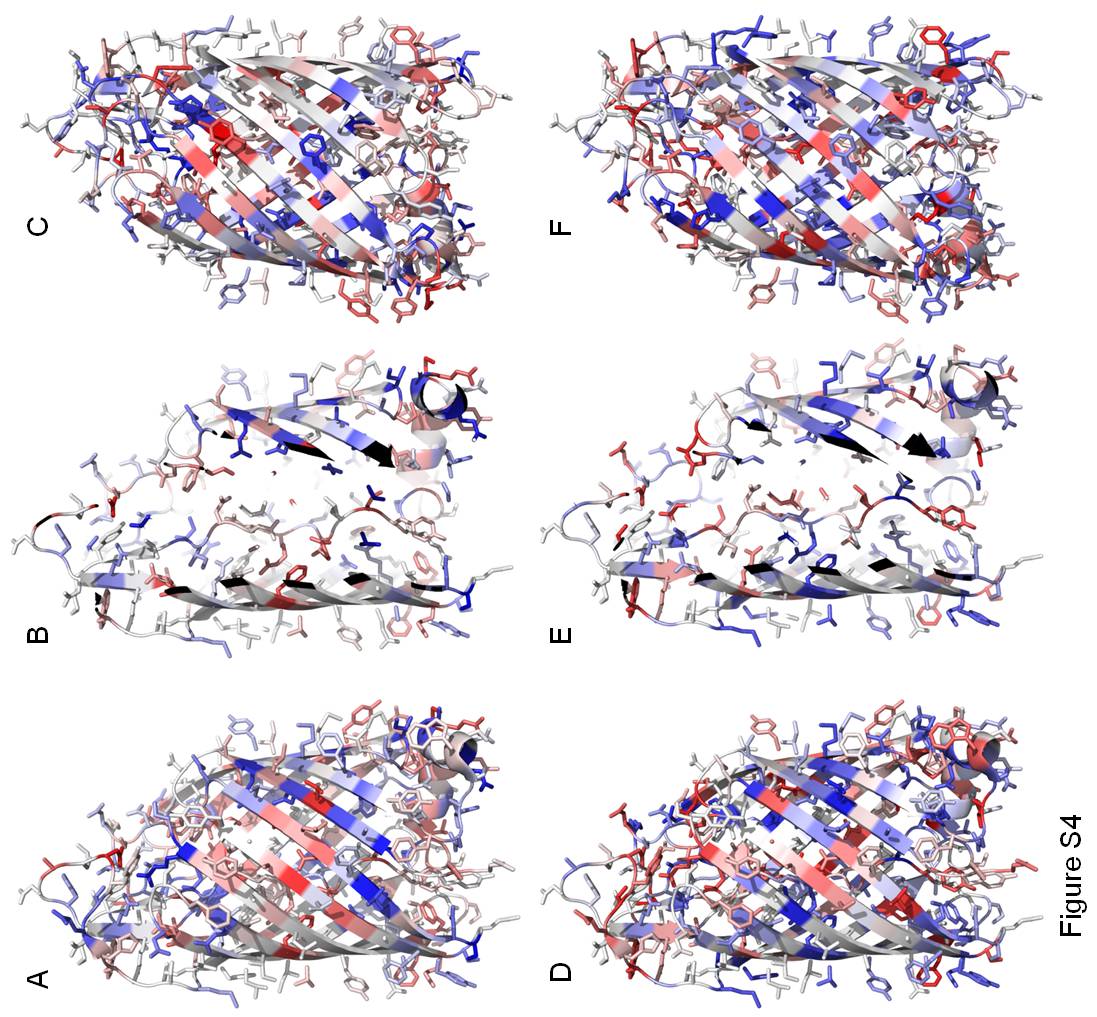}}
	\label{fig:Patterns_FigS4}
\end{figure}

\begin{figure}[h]
	\centering
		\centerline{\includegraphics[angle=270,origin=c]{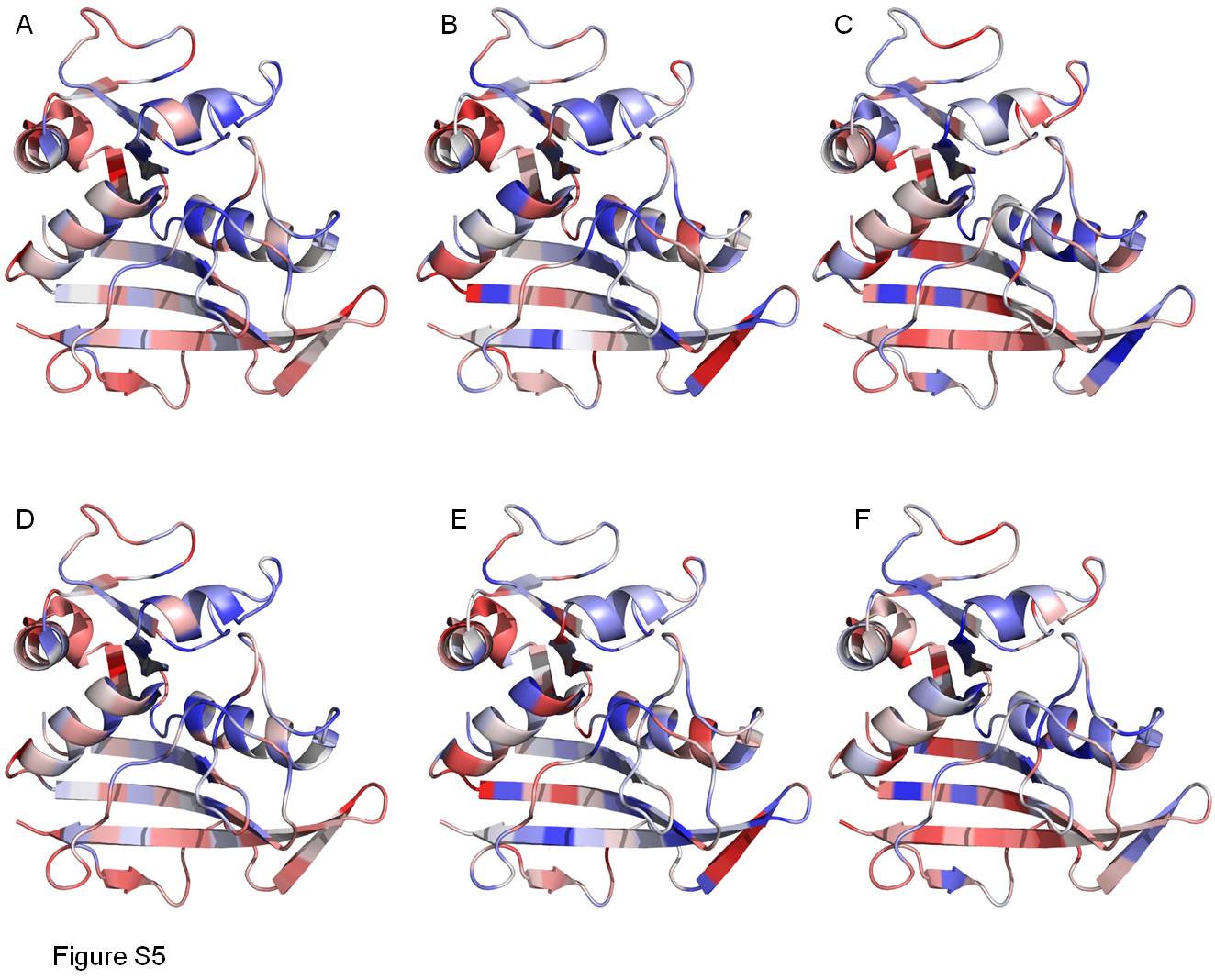}}
	\label{fig:Patterns_FigS5}
\end{figure}

\begin{figure}[h]
	\centering
		\centerline{\includegraphics[ angle=270,origin=c]{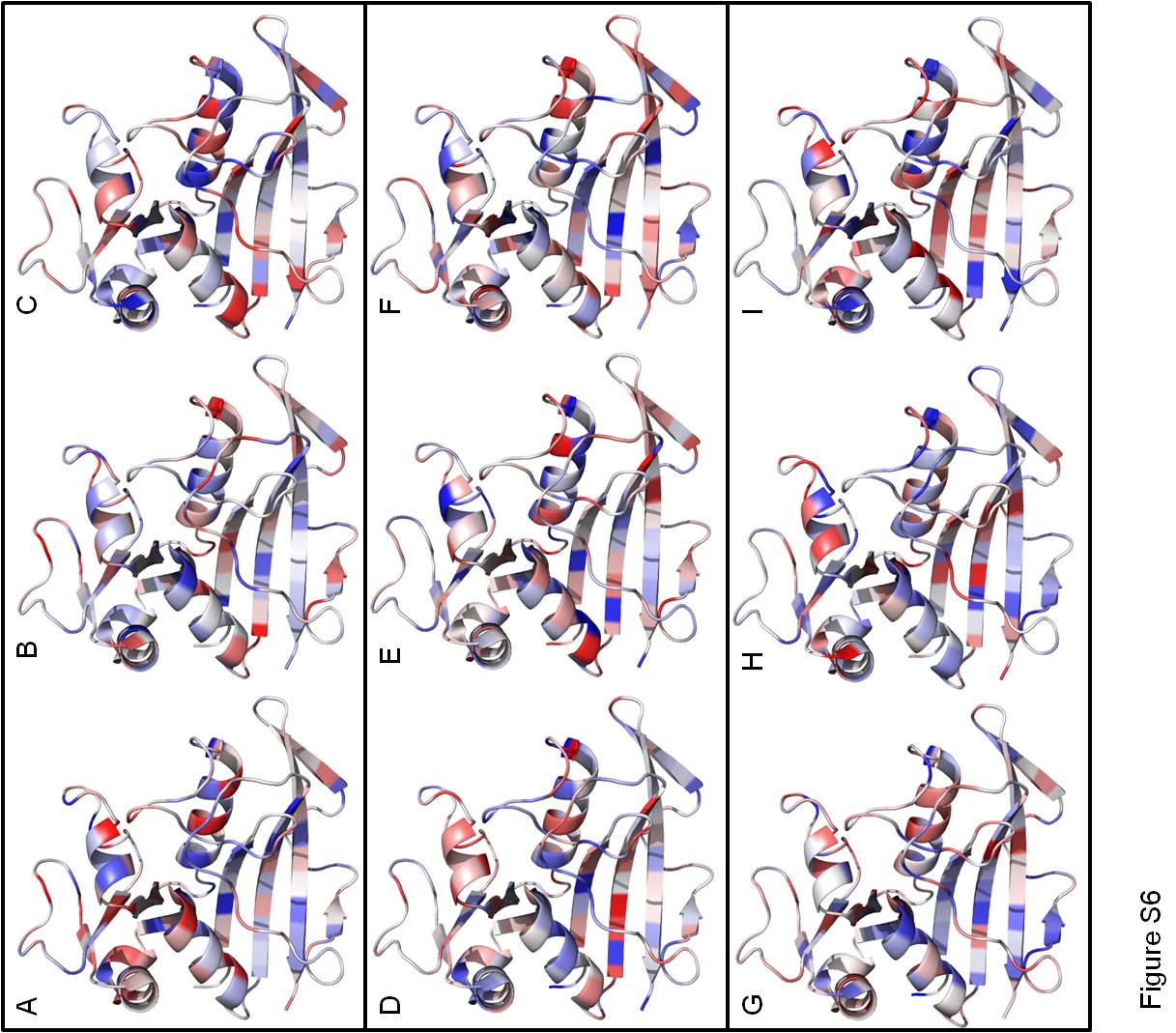}}
	\label{fig:Patterns_FigS6}
\end{figure}

\begin{figure}[h]
	\centering
		\centerline{\includegraphics[height=0.8\textheight,angle=270,origin=c]{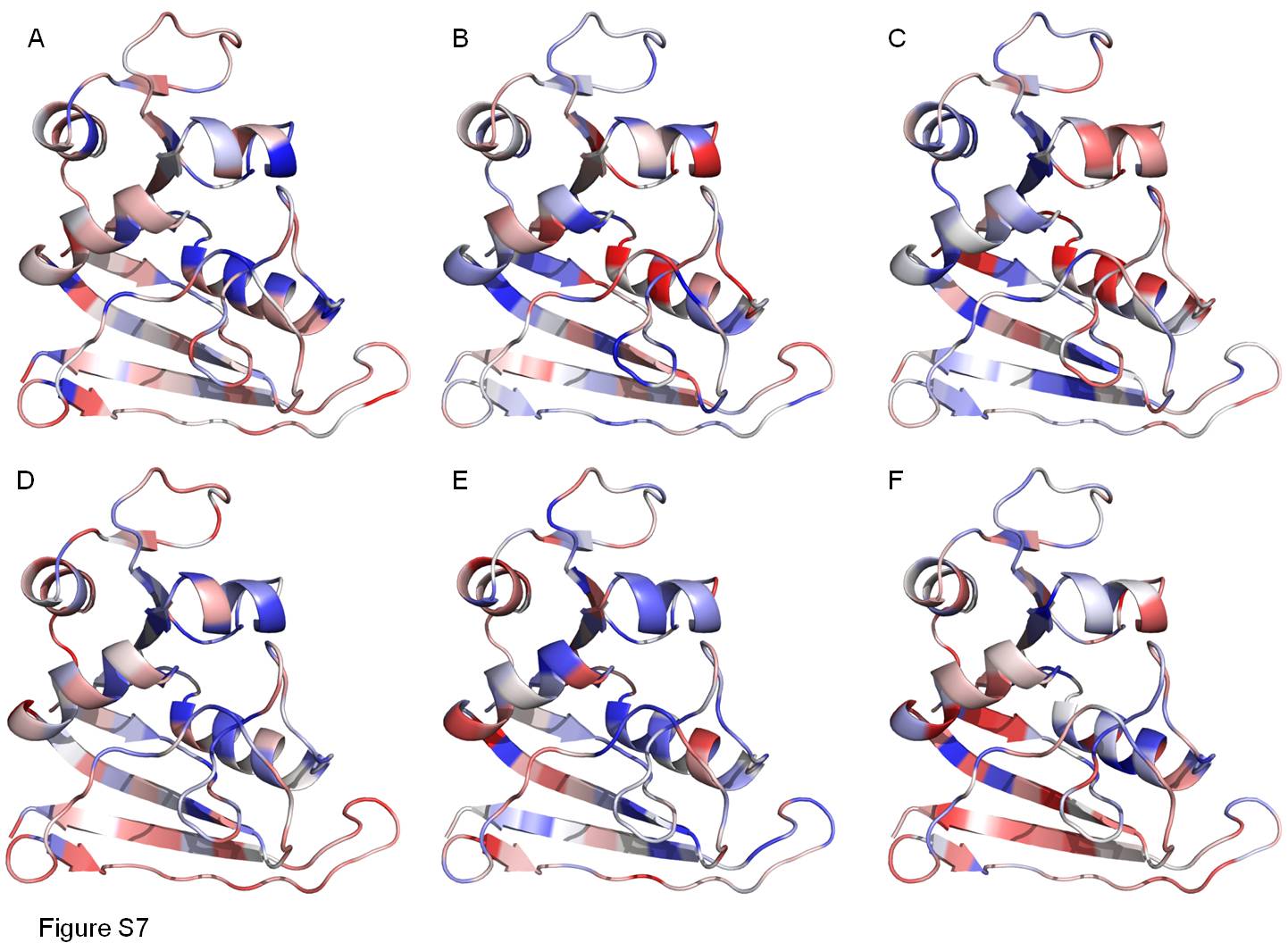}}
	\label{fig:Patterns_FigS7}
\end{figure}

\begin{figure}[h]
	\centering
		\centerline{\includegraphics[angle=270,origin=c]{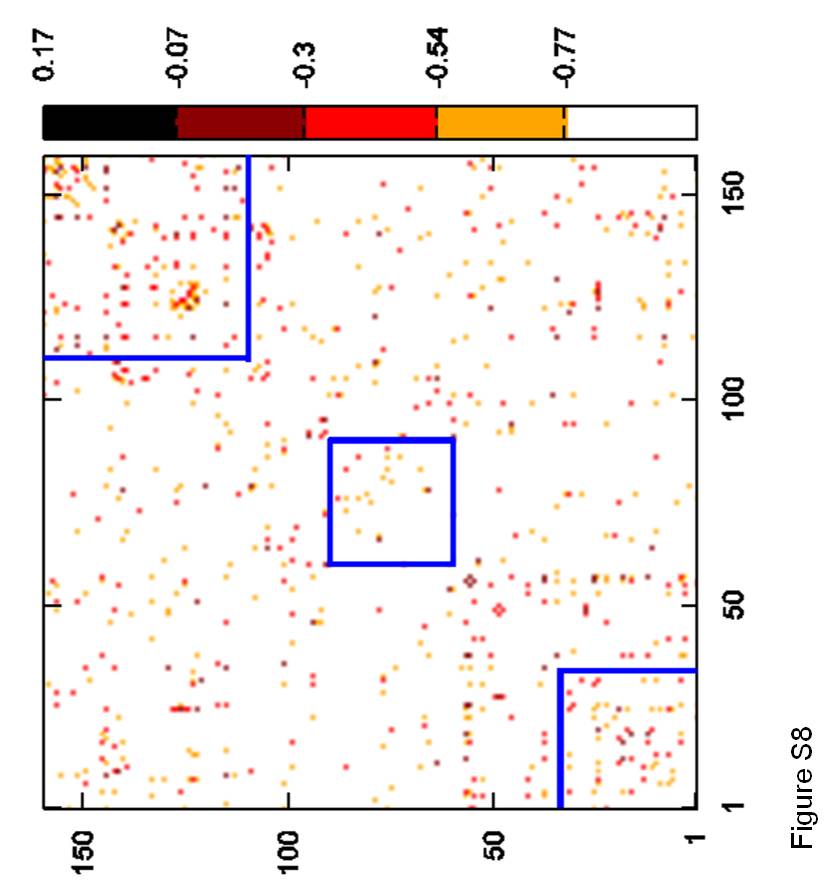}}
	\label{fig:Patterns_FigS8}
\end{figure}